\documentclass[aps,pra,twocolumn,showpacs,preprintnumbers,amsmath,amssymb,longbibliography]{revtex4-2}
\usepackage{graphicx}
\usepackage[usenames,dvipsnames]{xcolor}
\usepackage{amsmath}
\usepackage{cleveref}
\usepackage{makecell}
\usepackage{float}
\usepackage{lineno}
\usepackage[caption=false]{subfig}
\newcommand{\Tr}{\textrm{Tr}}

\newcommand{\bra}[1]{\ensuremath{\langle #1 |}}
\newcommand{\ket}[1]{\ensuremath{| #1 \rangle}}

\usepackage{xcolor}

\usepackage{ulem}
\usepackage{amsmath,amssymb}

\begin{document}
\title{Quantum simulation of spin-1/2 XYZ model using solid-state spin centers}

\author{Troy Losey$^1$}
\author{Denis R. Candido$^2$}
\author{Jin Zhang$^3$}
\email{jzhang91@cqu.edu.cn}
\author{Y. Meurice$^2$}
\author{M. E. Flatt\'{e}$^2$}
\author{S.-W. Tsai$^1$}
\affiliation{$^1$ Department of Physics and Astronomy, University of California, Riverside, California 92521, USA}
\affiliation{$^2$ Department of Physics and Astronomy, University of Iowa, Iowa City, Iowa 52242, USA}
\affiliation{$^3$ Department of Physics and Chongqing Key Laboratory for Strongly Coupled Physics, Chongqing University, Chongqing 401331, China}
\definecolor{burnt}{cmyk}{0.2,0.8,1,0}
\def\lt{\lambda ^t}
\def\note{note}
\def\beq{\begin{equation}}
\def\enq{\end{equation}}

\date{\today}
\begin{abstract}
In this work we propose a novel solid-state platform for creating quantum simulators based on implanted spin centers in semiconductors. We show that under the presence of an external magnetic field, an array of $S=1$ spin centers interacting through magnetic dipole-dipole interaction can be mapped into an effective spin-half system equivalent to the XYZ model in an external magnetic field. Interestingly, this system presents a wide range of quantum phases and critical behaviors that can be controlled via magnetic field and orientational arrangement
of the spin centers. We demonstrate our interacting spin chain can be tuned to both isotropic Heisenberg model and transverse-field Ising universality class. Notably, our model contains a line where the system is in a critical floating phase that terminates at Berezinskii–Kosterlitz–Thouless and Pokrovsky-Talapov transition points. We propose this system as the first solid-state quantum simulator for the floating phase based on spin centers.
\end{abstract}

\maketitle

\section{Introduction}\label{sec:introduction}

Quantum simulators designed for handling complex problems not solvable with classical computers have been a rapidly expanding field of quantum information science~\cite{PRXQuantum.2.017003,RevModPhys.86.153,Cirac2012}. Critical behaviors of complex systems may be investigated and tested with specially designed quantum simulators that contain the essential physics, and can be created and probed in the laboratory in a controlled way.  Critical phenomena are universal across distinct systems within the same universality classes, as they depend only on the symmetries and dimensionality of the system, and accordingly provide unifying principles that apply across very different fields of physics~\cite{sachdev_2011}. Quantum spin chains have been extensively studied due to their relative simplicity and rich critical behaviors~\cite{Farnell&Parkinson2010}, and can be exploited as quantum simulators. The physics of spin-half chains is particularly interesting and can be directly mapped to systems of fermions~\cite{Jordan&Wigner1928}, and moreover, effective spin models with spin-$S$ were proposed as quantum simulators for lattice field theories~\cite{PhysRevA.90.063603,PhysRevA.96.023603,PhysRevD.92.076003,PhysRevLett.121.223201,PhysRevD.98.094511}, with special interest in the $S=1$ truncation~\cite{PhysRevB.103.245137,PhysRevB.104.205112}. While neutral atoms, trapped ions, cavity arrays, quantum dots, superconducting circuits, photons, and nuclear spins have been been further studied as quantum simulators, issues with scalability or controlling and measuring individual qubits remain~\cite{PRXQuantum.2.017003,RevModPhys.86.153}. Our proposed spin chain quantum simulator is able to engineer numerous special Hamiltonian terms in order to simulate a wide variety of critical behavior in a single system and has pursuable avenues to address the issues faced by other quantum simulators.

\begin{figure*}[t!]
\begin{center}
\includegraphics[width=\textwidth]{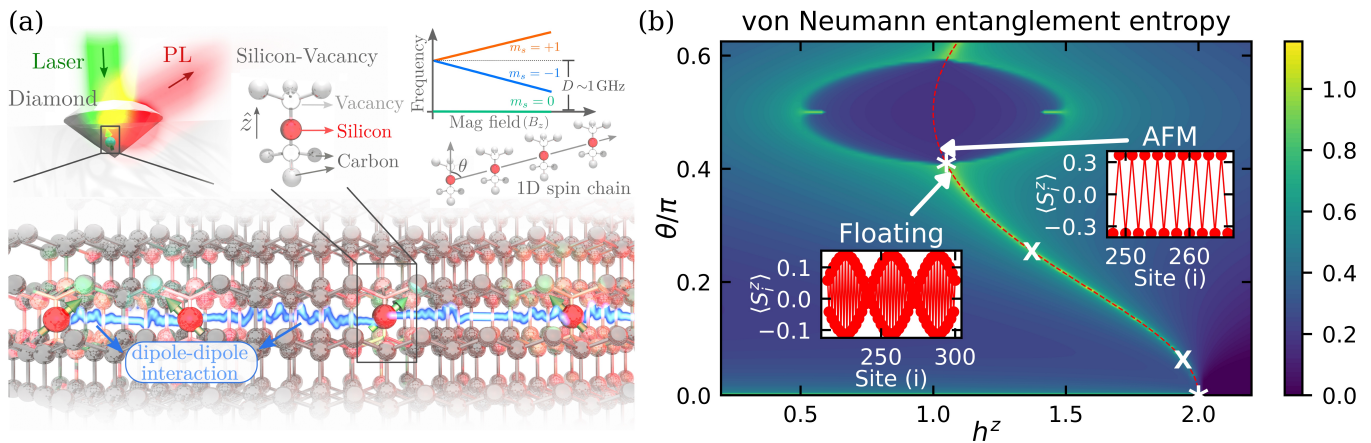}
\caption{\label{fig:NVphasediagramspinprofile}(a) Schematic representation of Silicon-Vacancy in Diamond and the corresponding spin array of SiV's coupled through dipole-dipole interaction. The spin centers are initialized by a laser, and optically read through the photoluminescence (PL). Also depicts the energy levels and z-axis from Eq.~\eqref{Hspincenter}, as well as the angle $\theta$ between the z-axis and the displacement vector between spin centers. (b) The phase diagram of the effective spin-1/2 SiV chain Hamiltonian in Eq.~\eqref{eq:rotatedHam}, demonstrated by the filled contour plot of von Neumann entanglement entropy $S_{\rm{vN}}$. The phase boundaries are indicated by the peaks of $S_{\rm{vN}}$. The two white X's and the two white stars show the locations of BKT and PT points, respectively. The ring is an Ising critical line, and the bright yellow curve connecting the PT point on the ring and the PT point at $\theta=0$ and $h^z=2$ is demonstrated to be the $\Gamma_x=0$ (red dashed) line, given by Eq.~(\ref{eq11}), which stays in a critical floating phase near the PT points and transitions via BKT points to a gapped phase for $0.06\pi \lesssim \theta \lesssim \pi/4$. The insets on the $\Gamma_x = 0$ line show the spin density profiles for the AFM phase inside the ring at $\theta=0.41\pi$ and the floating phase at $\theta=0.4\pi$. The $\theta=0$ line shows the critical partially magnetized phase with a high entanglement plateau for $h^z < 2$.}
\end{center}
\end{figure*}

Recently, defects with spin in solids (spin centers) have been demonstrated to be a promising platform for quantum information science due to their many applications~\cite{Awschalom2002,doi:10.1126/sciadv.abm5912,Schirhagl2014,Awschalom2018,PRXQuantum.2.017003,candido2020predicted}. These spin-$S$ centers are qudits that can be optically initialized with laser and optically read via the photoluminescence (PL) [Fig.~\ref{fig:NVphasediagramspinprofile}(a)], and present long spin coherence time even at room temperature. Due to the sensitivity of their energy levels to both magnetic and electric fields, they are also great candidates for quantum sensing and metrology~\cite{taylor2008high,koenraad2011single,electric-magnetic1,dolde2014,schirhagl2014nitrogen,van2015nanometre,degen2017quantum,flebus2018quantum,casola2018probing,electricnoise3,PhysRevX.10.011003,lee2020nanoscale,rustagi2020,RevModPhys.92.015004,magneticnoise7,PhysRevLett.121.023601,davis2021probing,dwyer2021probing,candido2021theory1}. Examples of solid-state spin centers are the negatively-charged nitrogen-vacancy (NV$^-$) \cite{Schirhagl2014,Awschalom2018,DOHERTY20131} and neutral silicon-vacancy (SiV$^0$) \cite{PhysRevB.77.245205,PhysRevLett.119.096402,doi:10.1126/science.aao0290,PhysRevLett.113.263602} spin centers in diamond; and divacancy spin centers in silicon carbide (SiC) \cite{koehl2011room,Seo2016,doi:10.1126/science.aax9406,PRXQuantum.2.040310}. Recent advances on the spatial implantation precision of spin centers~\cite{Fuechsle2012,doi:10.1063/1.4904909,doi:10.1021/acs.nanolett.5b05304} allow for the corresponding creation of room-temperature coherent spin arrays~\cite{Spinicelli_2011,doi:10.1021/nl102066q,Toyli2010,PhysRevApplied.12.064005}. Interestingly, the interaction between these spins can be set by the relative position of the spin centers within the crystal, and further tuned by applied external magnetic and electric fields. Moreover, as crystal hosts are much larger than the typical spin-spin implantation separation, scalability of these spin arrays appears encouraging. In addition, new avenues for work with interacting spin centers have been opened by studies and realizations of many-body phenomena in crystals containing an ensemble of interacting spin centers, e.g., discrete time-crystals~\cite{Choi2017,doi:10.1126/science.abk0603}, critical thermalization~\cite{PhysRevLett.121.023601}, Floquet prethermalization in a long-range spin interacting system~\cite{PhysRevLett.131.130401}, emergent hydrodynamics~\cite{Zu2021}, quantum metrology with strongly interacting spins~\cite{PhysRevX.10.031003} and Hamiltonian engineering via periodic pulse sequences~\cite{PhysRevX.10.031002}. Therefore, spin arrays made of spin centers are a promising solid-state candidate for the implementation of quantum simulators.

In this paper we exploit the remarkable recent advances in the creation and control of spin defects, and propose a solid-state quantum simulator for a spin-half system composed of an one-dimensional (1d) array of spin defects [See Fig.~\ref{fig:NVphasediagramspinprofile}(a)]. We show that this system presents various quantum phases and critical behaviors, which can be achieved using different spatial arrangements of spin centers and different values of the applied magnetic field [See Fig.~\ref{fig:NVphasediagramspinprofile}(b)]. More specifically, we consider an external magnetic field is applied to a 1d chain of anisotropic $S=1$ spin centers interacting via the magnetic dipole-dipole interaction. We show this creates an effective $S=1/2$ system with various spin-spin interaction terms. The corresponding phase diagram is characterized by calculation of the von Neumann entanglement entropy, and contains gapped magnetic orders and critical lines that correspond to the isotropic Heisenberg model, floating phases \cite{PhysRevB.19.2457,VillainBak1981,PhysRevLett.49.793,PBak_1982,PhysRevLett.122.017205} ending at Berezinskii-Kosterlitz-Thouless (BKT) \cite{Berezinsky:1970fr,Kosterlitz_1973} and Pokrovsky-Talapov (PT) \cite{PhysRevLett.42.65} points, and transverse-field Ising transitions. The presence of the floating phase makes our quantum simulator a promising potential candidate to study emerging incommensurate order and associated critical behavior described by Tomonaga-Luttinger liquid (TLL) theory \cite{F.D.M.Haldane_1981}.

The paper is organized as follows. In Section \ref{sec:model}, we start by introducing the general Hamiltonian that describes spin centers, followed by a discussion of the dipole-dipole coupling, and of the projection of the Hamiltonian onto an effective spin-half subspace. We then consider material candidates where this quantum simulator can be realized. In Section \ref{sec:results}, we present our results and discuss the most interesting regions of the phase diagram. We analyze the critical lines of the phase diagram in Sec.~\ref{subsec:theta0} for isotropic Heisenberg chain with an external field in the $z$ direction, Sec.~\ref{subsec:isingcircle} for Ising transitions, and Sec.~\ref{subsec:gammax0} for the critical floating phase. We summarize our results and discuss future work in Sec.~\ref{sec:conclusion}.

\section{Solid-state spin center implementation for quantum simulators} \label{sec:model}

In this section we establish a novel way for implementing quantum simulators in solid-state platforms. Our new proposal relies on the use of the spin center in semiconductors for creating interacting spin chain models in solid-state systems. Accordingly, we first provide the Hamiltonian for the spin centers and the dipole-dipole interaction between them. We then show that the total Hamiltonian can be mapped into a spin--1/2 XYZ chain with an applied magnetic field in the $x$--$z$ plane. Finally, we  discuss different challenges for experimental implementation with some solutions.

\subsection{Spin center Hamiltonian}

The Hamiltonian for the ground state of $S=1$ spin centers in solids can be generally described by a zero-field splitting term, plus the Zeeman interaction. For highly-symmetric spin centers, e.g., NV$^-$ and SiV$^0$ centers in diamond, and (hh) and (kk) di-vacancies in SiC, the low-energy effective Hamiltonian is described as arising primarily from two interacting electrons forming a triplet manifold $\{\ket{0},\ket{-1},\ket{1}\}$ with ground state Hamiltonian~\cite{doherty2011negatively,Schirhagl2014,PRXQuantum.2.040310,PhysRevB.77.245205,Seo2016,doi:10.1126/science.aax9406}
\begin{eqnarray}
    {\cal H}_{S} = h D \left(S^z\right)^2  + \frac{h \gamma}{2\pi} {\mathbf B}({\mathbf r}) \cdot {\mathbf S},
    \label{Hspincenter}
\end{eqnarray}
where $D$ is the zero-field splitting between the triplet states $m=0$ ($\ket{0}$) and $m = \pm 1$ ($\ket{\pm1}$); $S^x$,
$S^y$, and $S^z$ are the triplet spin--1 matrices, ${\gamma}$ is the gyromagnetic ratio (or g-factor), and $\mathbf{B}(\mathbf{r})$ the magnetic field at the position of the spin center, $\mathbf{r}$. Although spin centers also possess excited states, those are separated by relatively higher energy ($\sim$100~THz) compared to the energy scale associated to the ground state ($\sim$1~GHz), and therefore, will be neglected in this work. Accordingly, in the presence of an external magnetic field along the defect main symmetry axis, defined here as the $z$-direction ($\mathbf{B}(\textbf{r})=B\hat{z}$), the triplet energy levels read $ E_{m=\pm1}= hD \pm \frac{h\gamma}{2\pi}B$ and $E_{m=0} = 0$, and are illustrated in  Fig.~\ref{fig:NVphasediagramspinprofile}(a). For the majority of spin centers, $D\sim1$~GHz and ${\gamma/2\pi\sim 28}$~GHz/T, thus showing that spin centers can be easily manipulated with microwave frequency, and also respond sensitively to an external magnetic field.

\subsection{Magnetic dipole-dipole coupling between spin centers} \label{subsec:twoNV}

If we now consider an array of spin centers separated by inter-atomic distances, different spin centers will be coupled to each other through both magnetic dipole-dipole coupling, and exchange interaction. While the exchange interaction dominates for inter-atomic distances  $d\lesssim1$~nm~\cite{Kortan2016,PhysRevB.94.245424}, the dipole-dipole dominates for $d\gtrsim5$~nm. Assuming a chain with spin centers separated by distances $\gtrsim 10$~nm, we can disregard the exchange interaction, yielding the effective interacting Hamiltonian between spin centers $i$ and $j$,
\begin{eqnarray}
    {\cal H}_{\rm{int}}^{ij} = \frac{\mu_0(h\gamma/2\pi)^2}{4\pi|\mathbf{r}_{ij}|^3} [3(\mathbf{S}_j \cdot \hat{\mathbf{r}}_{ij})(\mathbf{S}_i \cdot \hat{\mathbf{r}}_{ij}) - (\mathbf{S}_j \cdot \mathbf{S}_i)],
    \label{dipoleij}
\end{eqnarray}
where $\mu_0$ is the vacuum permeability, $\mathbf{r}_{ij} = {\mathbf{r}}_{i} - {\mathbf{r}}_{j}$ is the displacement vector between spins $i$ and $j$ located at ${\mathbf{r}}_{i}$ and ${\mathbf{r}}_{j}$ respectively, and $\hat{\mathbf{r}}_{ij}={\mathbf{r}}_{ij}/|{\mathbf{r}}_{ij}|$.

In this paper, the $z$-axis is defined by the orientation of each spin center, predefined by the zero-field splitting term proportional to $(S^z)^2$ within Eq.~(\ref{Hspincenter}). Here, we express $\mathbf{r}_{ij}$ in spherical coordinates with corresponding angles $\theta$ and $\phi$. Since the $z$-axes of each spin center are aligned, our system possesses azimuthal symmetry and we set $\phi = 0$. With these assumptions, we can conveniently rewrite the magnetic dipole-dipole interaction between spin centers $i$ and $j$ [Eq.~(\ref{dipoleij})] as
\begin{eqnarray}
{\cal H}_{\rm{int}}^{ij} = \frac{\mu_0(h\gamma/2\pi)^2}{4\pi|\mathbf{r}_{ij}|^3}
\begin{bmatrix}
    S^x_i & S^y_i & S^z_i
\end{bmatrix}
\cdot \mathbf{T} \cdot
\begin{bmatrix}
    S^x_j \\ S^y_j \\ S^z_j
\end{bmatrix},
\end{eqnarray}
with dipole-dipole tensor
\begin{eqnarray}
\mathbf{T} =
\begin{bmatrix}
    3\sin^2(\theta) - 1 & 0 & \frac{3}{2}\sin(2\theta) \\
    0 & -1 & 0 \\
    \frac{3}{2}\sin(2\theta) & 0 & 3\cos^2(\theta) - 1
\end{bmatrix}.
\end{eqnarray}

Assuming typical electronic gyromagnetic ratio (${\gamma/2\pi\sim 28}$~GHz/T) and $|\mathbf{r}_{ij}|\sim 10$~nm, we obtain ${\cal H}_{\rm{int}}^{ij}/h \sim 50$~kHz, which is shown to be much stronger than the dephasing and relaxation rates of spin centers ($100$~$\mu$s to seconds)~\cite{Abobeih2018,doi:10.1126/sciadv.abm5912}.

Considering a spin chain of equally spaced ($|\mathbf{r}_{ij}|=|\mathbf{r}|$) $N$ spin centers oriented along a straight line with a polar angle $\theta$ [See Fig.~\ref{fig:NVphasediagramspinprofile}(a)], and assuming only nearest-neighbor (NN) interactions due to the short-range character of the dipole-dipole interaction, we obtain the total Hamiltonian
\begin{eqnarray}
    {\cal H}_{tot} = \sum_{i}{\cal H}_{S_i} + \sum_{i,j}{\cal H}_{\rm{int}}^{ij}\delta_{i,j\pm1}.
    \label{H_tot}
\end{eqnarray}
We thus have an interacting spin chain that can simulate nontrivial physics of strongly correlated systems. Notice that if the dipole-dipole coupling is strong, long-range interaction should be kept and novel quantum phenomena like spontaneous breaking of continuous symmetries in low dimensions can be observed \cite{PhysRevLett.119.023001,ChenChengDipolarXY2023}.

\subsection{Effective spin-half Hamiltonian}
\label{subsec:spinhalfH}
Despite the spin--1 character of our spin centers, we can effectively map the total spin--1 interacting Hamiltonian [Eq.~(\ref{H_tot})] into a spin--1/2 interacting Hamiltonian. To do so, we first apply a magnetic field $\textbf{B}=B_c \hat{z}$ with  $B_c\sim D/(\gamma/2\pi)$, such that the levels $\ket{-1}$ and $\ket{0}$ are nearly degenerate. Under this condition, the state $\ket{1}$ is separated from both $\ket{-1}$ and $\ket{0}$ by $\sim 1$~GHz. As the coupling between different spin centers is $\sim 50$~kHz, non-degenerate perturbation theory guarantees that the effect of level $\ket{1}$ within the manifold spanned by ${\ket{-1}}$ and $\ket{0}$ can be neglected. Accordingly, by projecting the total spin--1 interacting Hamiltonian [Eq.~(\ref{H_tot})] onto the low energy $\{\ket{-1},\ket{0}\}$ subspace, we obtain the spin--1/2 interacting Hamiltonian [See Appendix~\ref{appdx:effspinhalf}]
\begin{widetext} 
\begin{eqnarray}
\label{eq:effectiveham}
 H &=& J\sum_{i=1}^{N-1} \bigg\{ \left[3\sin^2(\theta) - 1\right]\sigma_i^x \sigma_{i+1}^x - \sigma^y_i \sigma^y_{i+1}  + \frac{3\cos^2(\theta) - 1}{2} \sigma^z_i \sigma^z_{i+1} + \frac{3\sin(2\theta)}{2\sqrt{2}}\left(\sigma^x_i \sigma^z_{i+1} + \sigma^z_i \sigma^x_{i+1}\right)\bigg\} \nonumber \\ &-& J\sum_{i=1}^{N}\left\{\left[h^z +3\cos^2(\theta) - 1\right]\sigma^z_i + \frac{3\sin(2\theta)}{\sqrt{2}}\sigma^x_i \right\},
\label{H-spin1/2}
\end{eqnarray}
\end{widetext}
with $J={\mu_0 (h\gamma/2\pi)^2}/{8\pi|\mathbf{r}|^3}$, $h^z=(E_{m=-1}-E_{m=0}) / J$
and Pauli matrices $\sigma_{x,y,z}$ defined as $\sigma^{+}=(\sigma^x+i\sigma^y)/2=\ket{0}\bra{-1} $. From now on, we will set $J=1$.

The Hamiltonian in Eq.~(\ref{H-spin1/2}) is invariant under the transformations $\theta\rightarrow\theta+\pi$, or $\theta\rightarrow \pi - \theta$ combined with a $\pi$-rotation of all spins around the $z$-axis, thus we restrict our analysis to $\theta \in [0, \pi/2]$ in the following calculations. We set $A = 3\sin^2(\theta)-1$, $B = [3\cos^2(\theta)-1]/2$, and $C = 3\sin(2\theta)/(2\sqrt{2})$; then rotate the spins around the $y$-axis by an angle $\alpha$ [See Appendix~\ref{appd2}]. Choosing proper values of $\alpha$, we eliminate the $\sigma^x_{i}\sigma^z_{i+1}+\sigma^z_{i}\sigma^x_{i+1}$ terms, yielding the following XYZ model with an effective external field in the $x$--$z$ plane
\begin{eqnarray}
\label{eq:rotatedHam}
\nonumber \tilde{H} &=& \sum_{i=1}^{N-1} \left( J_x \sigma^x_i \sigma^x_{i+1} + J_z \sigma^z_{i} \sigma^z_{i+1} -\sigma^y_i \sigma^y_{i+1}\right) \\ &-& \sum_{i=1}^{N} \left( \Gamma_z\sigma^z_i + \Gamma_x\sigma^x_i \right),
\end{eqnarray}
and parameters
\begin{eqnarray}
&& J_{x(z)} = \mp \sqrt{\frac{(A-B)^2}{4}+C^2} + \frac{A+B}{2}, \label{eq9} \\
&& \Gamma_z = \left(h^z+2B\right)\cos(\alpha) + 2C\sin(\alpha) \label{eq10},\\
&& \Gamma_x = -\left(h^z+2B\right)\sin(\alpha) + 2C\cos(\alpha) \label{eq11}, \\
&& \label{eq:alphavstheta} \alpha = 
\begin{cases}
    \frac{1}{2}\arctan\left(\frac{2C}{B-A}\right), & \text{if } 0 \leq \theta < \arcsin(\frac{2}{3}) \\
    \frac{\pi}{4},  & \text{if } \theta = \arcsin(\frac{2}{3}) \\
    \frac{1}{2}\arctan\left(\frac{2C}{B-A}\right) + \frac{\pi}{2}.  & \text{if } \arcsin(\frac{2}{3}) 
 < \theta \leq \frac{\pi}{2}
\end{cases} \label{eq12}
\end{eqnarray}
It is noticed that the same model can also be engineered using $p$-orbital bosons in optical lattices \cite{XYZpOrbital2013} and the long-range dipolar XYZ model in one or two dimensions can be built in artificial arrays of Rydberg atoms in optical tweezers \cite{PRXQuantum.3.020303}. 

\begin{figure}[h!]
  \centering
    \includegraphics[width=0.48\textwidth]{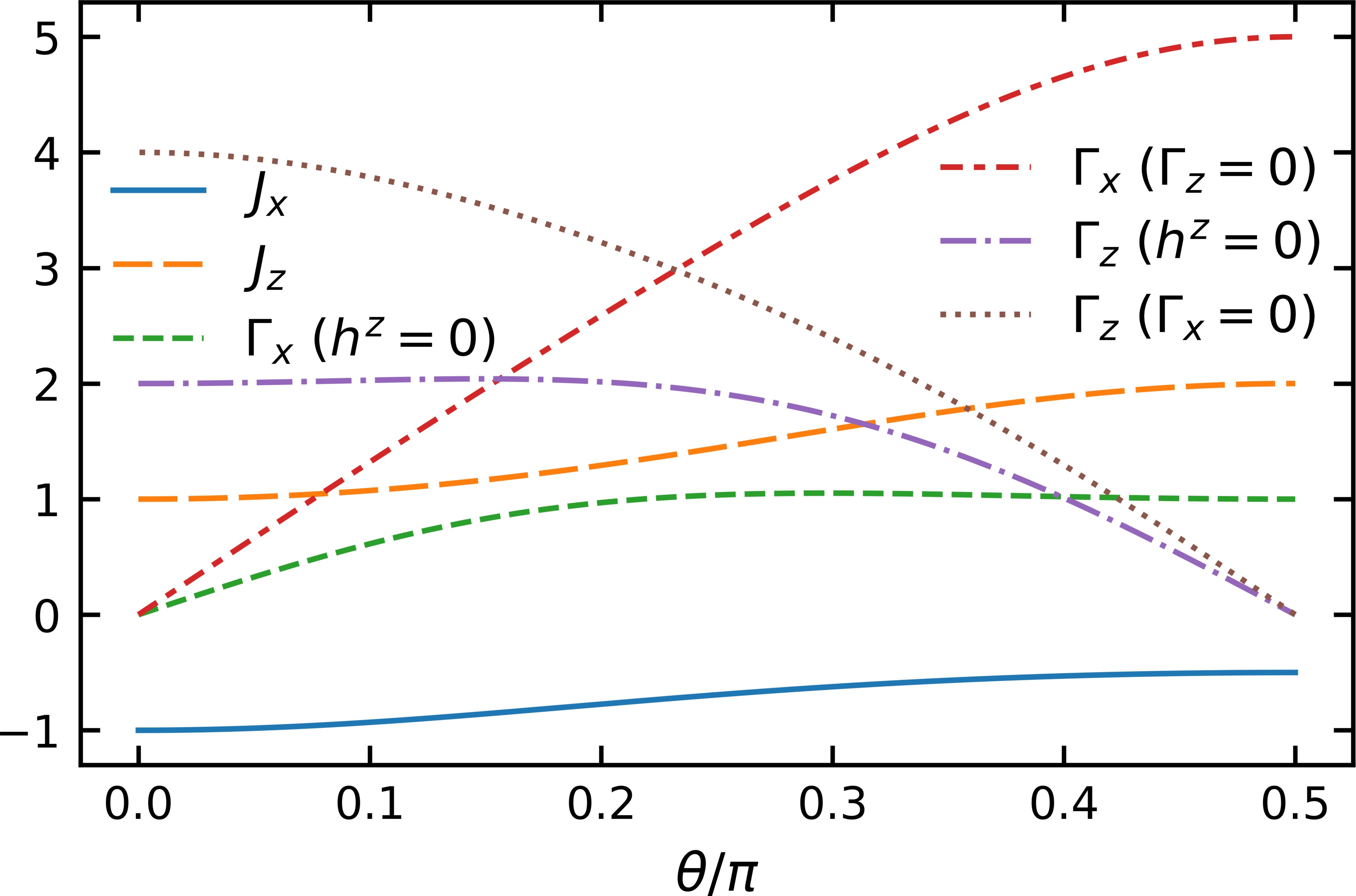}
    \caption{\label{fig:parametervstheta} The dependence of the coupling constants of the rotated Hamiltonian on $\theta$. The NN interaction strengths $J_{x(z)}$ only depend on $\theta$, while the rotated fields $\Gamma_{x(z)}$ depend on both $\theta$ and $h^z$. The values of $\Gamma_{x(z)}$ as a function of $\theta$ for fixed $h^z=0$ and fixed $\Gamma_{z(x)}=0$ are depicted here, respectively.
    }
\end{figure}

Without $\Gamma_{x(z)}$ fields, the XYZ model is integrable \cite{baxter2016exactly} and can be exactly solved using the Bethe ansatz \cite{CAO2014185}. However, the Hamiltonian [Eq.~\eqref{eq:rotatedHam}] is more complex due to the presence of external fields. As $J_{x(z)}$ and $\Gamma_{x(z)}$ depend on $\theta$ and $h^z$, analyzing the corresponding parameter space shows what quantum phases and critical phenomena our quantum simulator can explore. Figure~\ref{fig:parametervstheta} shows the effective Hamiltonian parameters [Eqs.~(\ref{eq9})-(\ref{eq12})] as functions of $\theta$. Notice that the interaction strengths $J_x$ and $J_z$ only depend on $\theta$, while the effective field components, $\Gamma_x$ and $\Gamma_z$ depend on both $\theta$ and $h^z$. At $\theta = 0$, we have $J_x = -1$, $J_z = 1$, $\Gamma_x = 0$ for arbitrary $h^z$, and $\Gamma_z$ linear in $h^z$. After a $\pi$-rotation around the $z$-axis for even or odd sites, the Hamiltonian [Eq.~(\ref{eq:rotatedHam})] becomes a Heisenberg model in an external field along the $z$-direction. Conversely, near $\theta = \pi/2$, $\alpha \approx \pi/2$, $\Gamma_z \approx 0$ and depends little on $h^z$, the $J_z$ coupling dominates the interactions, and the $\Gamma_x$ field can drive a continuous quantum phase transition between disordered and antiferromagnetic (AFM) phases as the transverse-field Ising model does \cite{R_B_Stinchcombe_1973}. As $\theta$ increases from $0$ to $\pi/2$, $J_x$ changes from $-1$ to $-0.5$, and $J_z$ changes from $1$ to $2$, while $|J_z| \geq 1 \geq |J_x|$ is always true. The interactions of the system favor ferromagnetic (FM) order in $x$--$y$ plane and AFM order in $z$-direction. Due to noncommutative properties of spin operators, the competitions between tendencies toward FM and AFM orders in different directions can result in nontrivial quantum criticalities \cite{Langari1998,Langari2004,PhysRevB.83.012402}.

Without $\Gamma_{x(z)}$ fields, the Hamiltonian has a set of $\mathbb{Z}_2$ symmetries and is invariant under a $\pi$ rotation about any of the $x$, $y$, or $z$ axes. The uniform fields $\Gamma_{x(z)}$ will orient all spins in the same direction and explicitly break the $\mathbb{Z}_2$ symmetry about the other two axes. There are two special lines corresponding to $\Gamma_x = 0$ and $\Gamma_z = 0$ respectively, where the Hamiltonian [Eq.~(\ref{eq:rotatedHam})] depends only on $\theta$. The rotated fields along the two lines read
\begin{eqnarray}
\label{eq:verticallines}
\Gamma_z &=& 2C/\sin(\alpha) \text{ for } \Gamma_x = 0, \\
\Gamma_x &=& 2C/\cos(\alpha) \text{ for } \Gamma_z = 0.
\end{eqnarray}
While increasing $\theta$ along the $\Gamma_z=0$ line, the Hamiltonian [Eq.~(\ref{eq:rotatedHam})] begins at $\theta = 0$ at an SU(2) symmetric Heisenberg point where $\Gamma_x=0$, then immediately becomes Ising-like, and ends at $\theta=\pi/2$ with $\Gamma_x=5$. As a result, only Ising transitions are possible on the $\Gamma_z=0$ line, and the positive increasing $\Gamma_x$ field will simply enhance FM order in the $x$--direction. We have explored this line numerically and found no critical behaviors beyond those at $\theta = 0$.

Along the $\Gamma_x=0$ line, the Hamiltonian [Eq.~(\ref{eq:rotatedHam})] begins at $\theta=0$ as a Heisenberg model with an external field $\Gamma_z = 4$ [See Eq.~(\ref{eq:hamtheta0})], which is a commensurate-incommensurate transition point [See Sec.~\ref{subsec:gammax0}], and ends at $\theta=\pi/2$ as an AFM phase with $\Gamma_z=0$, where $J_z=2$ dominates the interaction strength. Throughout the $\Gamma_x=0$ line, the field $\Gamma_z$ will attempt to align all spins, while the AFM coupling $J_z>0$ will attempt to anti-align NN spins in the $z$ direction. This competition may cause the proliferation of domain walls in the presence of $J_{x(y)}$ terms and nontrivial incommensurate order can appear \cite{VillainBak1981,PhysRevLett.49.793}. In summary, based on analysing the parameter space of the Hamiltonian [Eq.~(\ref{eq:rotatedHam})], we have discovered our quantum simulator can probe the Heisenberg model in an external field, Ising transition lines, and novel quantum criticalities associated with incommensurate orders. These rich features are confirmed and extensively studied in Sec.~\ref{sec:results} via large-scale numerical calculations.

\subsection{Experimental feasibility}

Here we explore and discuss the best spin center candidates for realizing our proposal, in addition to assessing the feasibility of it. As already mentioned, there are many examples for spin--1 spin centers in solid-state systems that are relevant for our proposal, including the different di-vacancies in SiC, and both NV$^-$ and SiV$^0$ centers in diamonds. Within these candidates, the one that is more established in terms of understanding, knowledge, control, implantation precision and manipulation is the NV$^-$ center in diamonds. However, the majority of the Nitrogen isotopes, $^{14}$N and $^{15}$N, have a nuclear spin. Since the spin of the NV$^-$ centers are close to the N atoms, the hyperfine interaction between them is substantial, with corresponding strength $\gtrsim 2$~MHz~\cite{DOHERTY20131,PhysRevB.94.155402,PhysRevB.100.075204}. As the magnetic dipole-dipole interaction produces NV-NV coupling around $\sim 50$~kHz, this interaction would be suppressed by the coupling of the NV$^-$ to the N nuclear spin. Due to this reason, the realization of our proposal will be optimal for solid-state systems that do not present strong hyperfine interaction between spin centers and non-zero spin nuclei. Accordingly, both di-vacancies in SiC, and SiV$^0$ centers in diamond are shown to be better candidates for our proposal as the majority of their atoms (Si and C) do not have nuclear spin. We emphasize that although $^{12}$C has zero nuclear spin and $^{13}$C does not, $^{13}$C only represents $\sim1\%$ of the whole Carbon atoms of the crystal, and accordingly, is not expected to suppress the dipole-dipole interaction.

Despite the rich theoretical predictions for the different phases of our system, it presents few experimental challenges. The spin center chain must be compact enough for the magnetic dipole-dipole interaction to be measurable.  NV$^-$ centers have been implanted with a separation of $16 \pm 5$ nm~\cite{doi:10.1021/nl504441m}. This is nearing the previously mentioned $\sim 10$ nm minimum spin center separation where the magnetic dipole-dipole interaction is maximized, while still dominating the exchange interaction. Additionally, the variation in the relative placement of the spin centers must be small enough to ensure that the couplings do not significantly change between different pairs of NN spin centers, since we assume the spin centers are equally spaced and in a straight line.

Another challenge is that the spin center couplings are proportional to $J$, which in temperature units reads $J/k_B\approx 2 ~\mu$K (considering spin centers separated by $10$~nm). This very low temperature regime is currently unachievable in experimental setups, thus setting practical limitations to our proposal. To overcome this, we propose to engineer stronger spin-spin coupling $J$ by taking advantage of interactions mediated by bosonic modes, e.g., photonic~\cite{Bernien2013,doi:10.1126/science.1253512,Hensen2015}, polaritonic~\cite{PhysRevApplied.10.024011}, phononic ~\cite{PhysRevLett.117.015502,PhysRevLett.120.213603} and magnonic~\cite{trifunovic2013long,flebus2019entangling,zou2020tuning,candido2020predicted,neumanprl2020,fukami2021,peraca2023quantum}. In particular, hybrid schemes with magnon-mediated spin-spin coupling were proposed~\cite{candido2020predicted,fukami2021}, and yield both easy scalability and strong coupling $J/h \approx 1$~MHz between NV's distanced by $\sim1$~$\mu$m. As a consequence, these schemes also relax the requirements for a compact chain of precisely placed spin centers mentioned above. In short, hybrid quantum systems offer an alternative solution for pushing the critical temperatures for realizing different floating phases using spin centers.

\section{Results}\label{sec:results}

In order to understand the phase diagram in the whole parameter space, we perform finite-size density-matrix renormalization group (DMRG) calculations \cite{PhysRevLett.69.2863, PhysRevB.48.10345,PhysRevLett.75.3537} [See Appendix~\ref{appdx:DMRG}]. Our DMRG calculations typically use an odd number of sites to avoid domain walls forming in the center of the chain due to finite-size boundary effects [See Appendix~\ref{appdx:EvenVsOdd}].
We utilize the von Neumann entanglement entropy as the universal phase-transition indicator and graph the phase diagram of our Hamiltonian [Eq.~\eqref{eq:rotatedHam}]. The von Neumann entanglement entropy $S_{\rm{vN}}$ is a measure of entanglement between the subsystem $\mathcal{A}$ of a quantum many-body system and its complement $\mathcal{B}$
\begin{eqnarray}
S_{\rm{vN}} = - \Tr[ \hat{\rho}_{\mathcal{A}} \ln(\hat{\rho}_{\mathcal{A}})],
\label{eq15}
\end{eqnarray} 
where $\hat{\rho}_{\mathcal{A}} = \Tr_{\mathcal{B}} \hat{\rho}$ is the reduced density matrix for the subsystem $\mathcal{A}$, and $\hat{\rho}$ is the density matrix of the whole system, which is equal to $\ket{\Psi_0}\bra{\Psi_0}$ if the system is in the ground state $\ket{\Psi_0}$. Here, we only consider the case where $\mathcal{A}$ is half of the system. At a critical point on phase transition lines or in gapless phases, conformal field theory (CFT) predicts that the entanglement entropy of a system with open boundary conditions diverges logarithmically with the system size as \cite{AffleckCritical1991,HOLZHEY1994443,VidalEntangle2003,PasqualeCalabrese_2004,Calabrese_2009}
\begin{eqnarray}
\label{eq:eescaling}
    S_{\rm{vN}} = s_0 + \frac{c}{6}\ln (N),
\end{eqnarray}
where $c$ is the central charge referred to as the conformal anomaly, and $s_0$ is a non-universal constant. This universal scaling behavior of $S_{\rm{vN}}$ can be used to detect critical points or lines, and calculate the central charge to determine the system's universality class.

Figure~\ref{fig:NVphasediagramspinprofile}(b) shows the phase diagram of the model in the $\theta$--$h^z$ plane using von Neumann entanglement entropy [Eq.~(\ref{eq15})]. As mentioned above in Sec.~\ref{subsec:spinhalfH}, the rotated Hamiltonian [Eq.~\eqref{eq:rotatedHam}] at $\theta = 0$ is a Heisenberg model in an external field along the $z$-direction $\Gamma_z = 2+h^z$. At $\Gamma_z=0$ or $h^z=-2$, our solid-state system will simulate the SU(2) Heisenberg model. According to the results in Refs.~\cite{Langari1998, PhysRev.150.321}, the model is in the critical partially magnetized phase for $\Gamma_z \in (0, 4)$ or $h^z \in (-2,2)$, and goes to the fully magnetized phase for $\Gamma_z > 4$ via a commensurate-incommensurate PT transition at $\Gamma_z = 4$ or $h^z = 2$. These behaviors are confirmed by the high-entanglement segment ending at $h^z=2$ on the bottom line of the phase diagram.

Close to $\theta = \pi/2$, the interaction strength between NN spin projections along the $z$-direction dominates, so the transverse field can drive an Ising transition with $\nu=1$ and $c=1/2$. In the phase diagram Fig.~\ref{fig:NVphasediagramspinprofile}(b), there is an Ising critical ring centered at $\theta=\pi/2$ and $h^z=1$. Inside this ring, our system is in an AFM phase manifesting non-zero staggered magnetization per site in the $z$-direction $m^z_{stag} = [\sum_i (-1)^i \langle S^z_i\rangle]/N$ [See inset of Fig.~\ref{fig:NVphasediagramspinprofile}(b)]. The system is in the disordered phase outside the ring, except for one bright yellow line where the entanglement entropy is large. This line coincides with the special line defined by $\Gamma_x = 0$ in Eq.~(\ref{eq11}) [See red dashed line in Fig.~\ref{fig:NVphasediagramspinprofile}(b)]. Large entanglement on the line indicates the existence of a critical phase that is demonstrated to be the floating phase [See Sec.~\ref{subsec:gammax0}]. The ends of this line are indicated by the asterisk symbol and correspond to PT transition points that are on the boundary of the critical floating phase [See inset Fig.~\ref{fig:NVphasediagramspinprofile}(b)]. Along this line, the floating phase emerges from the upper PT point down to $\theta \approx \pi/4$ (top X marker) and from the bottom PT point up to $\theta \approx 0.06\pi$ (bottom X marker). On the line for $0.06\pi \lesssim \theta \lesssim \pi/4$ (between the X markers), the system is a gapped phase, separated from the floating phase by the two BKT transitions which are represented by the X markers. In the following subsections, we analyze the properties of these quantum phases and phase transitions with substantial numerical evidence.

\subsection{$\theta = 0$: Heisenberg chain}
\label{subsec:theta0}

\begin{figure}[h!]
\centering
\includegraphics[width=.48\textwidth]{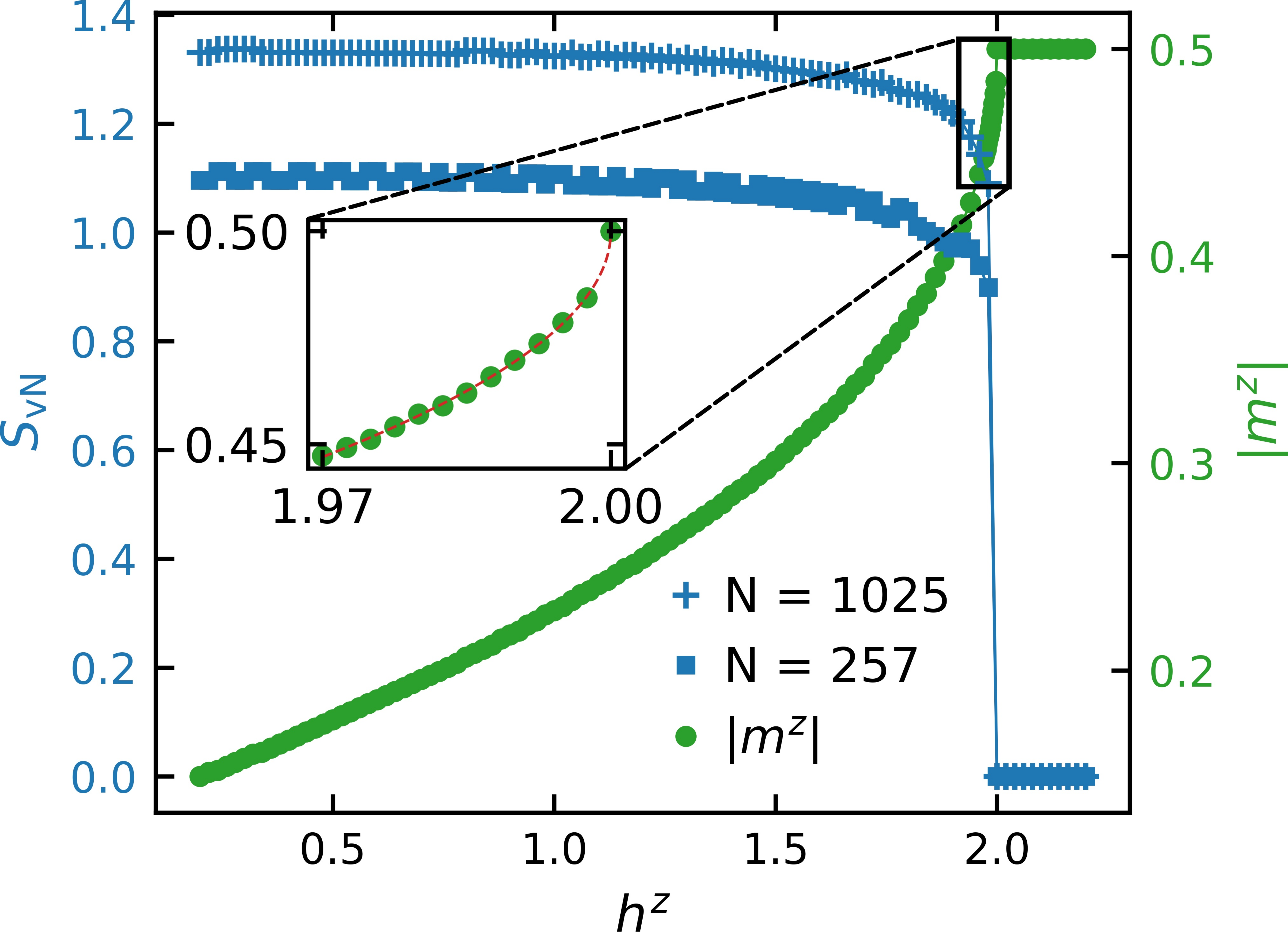}
\caption{\label{fig:s1mzvshztheta0}The entanglement entropy $S_{\rm{vN}}$ for $N=257$ and $1025$, and the magnetization per site $|m^z|$ for $N=1025$ as functions of $h^z$ at $\theta = 0$. The large entanglement entropy with small oscillations (blue pluses and blue squares) in the partially magnetized phase sharply drops to zero in the fully magnetized phase. The magnetization per site increases with $h^z$ in the partially magnetized phase and saturates at $1/2$ in the fully magnetized phase. The inset shows a fit of the power-law scaling of $|m^z|$ near the phase transition between the partially and fully magnetized phases.}
\end{figure}

The spin--$1/2$ chain Hamiltonian in Eq.~(\ref{eq:rotatedHam}) for $\theta=0$ reads
\begin{eqnarray}
\label{eq:hamtheta0}
\nonumber H_{\theta=0} &=& \sum_i \left( -\sigma^x_i \sigma^x_{i+1} - \sigma^y_i \sigma^y_{i+1} + \sigma^z_{i}\sigma^z_{i+1} \right) \\ &-& \left(h^z + 2\right)\sum_i \sigma^z_{i},
\end{eqnarray}
which is a Heisenberg model with an external field in the $z$-direction after a $\pi$-rotation around the $z$-axis for even or odd sites. It has an SU(2) Heisenberg point at $h^z = -2$, which is in the BKT universality class \cite{BANKS1976119} with a central charge $c = 1$. As we increase $h^z$ from $-2$, a non-zero magnetization per site in the $z$-direction, $m^z = \left[\sum_i \langle S^z_i\rangle\right]/N$, is induced [See Fig.~\ref{fig:s1mzvshztheta0}]. As the total magnetization is conserved, the magnetization in the $z$-direction $\sum_i \langle S^z_i\rangle$ increases in steps of $1/2$ and the magnetization per site is continuous in the large $N$ limit. Based on previous studies \cite{Langari1998,Langari2004}, before the magnetization saturates, the system is in the critical partially magnetized phase with logarithmically diverging entanglement entropy. Our data shows that large $S_{\rm{vN}}$ persists until $h^z=2$ where $S_{\rm{vN}}$ suddenly jumps to zero and the magnetization per site saturates at $1/2$. When the system size is increased from $N=257$ to $N=1025$, the increment in $S_{\rm{vN}}$ is almost a constant about $0.23$, consistent with CFT prediction $\ln(4)/6$ with central charge $c=1$. It has been argued that $S_{\rm{vN}}$ for Eq.~\eqref{eq:hamtheta0} has parity oscillations that depend on the Fermi momentum and Luttinger liquid parameter \cite{XavierParity2011}, which also results in oscillations in $S_{\rm{vN}}$ as a function of $h^z$ shown in Fig.~\ref{fig:s1mzvshztheta0}. The oscillation amplitude decreases with the system size and $S_{\rm{vN}}$ becomes smooth in the large $N$ limit. For $h^z \geq 2$, all the spins point along the $z$-direction and the system is in the fully magnetized phase with maximum magnetization.

In the thermodynamic limit, the magnetization per site along the $z$-direction $m^z$ changes continuously with $h^z$, thus there is a power-law scaling for $m^z$ near the phase transition inside the partially magnetized phase. In the inset of Fig.~\ref{fig:s1mzvshztheta0}, we plot the $m^z$ as a function of $h^z$ near $h^z_c=2$. Assuming $m^z \sim 1/2 - (h^z_c - h^z)^\beta$, we obtain $\beta = 0.501(1)$ from a curve fit. Our numerical results are consistent with the transition between the incommensurate partially magnetized phase and the commensurate fully magnetized phase belonging to the PT universality class with $\beta = \nu = 1/2$ \cite{PhysRevLett.42.65}.

\begin{figure*}[t!]
  \begin{center}
    \includegraphics[width=\textwidth]{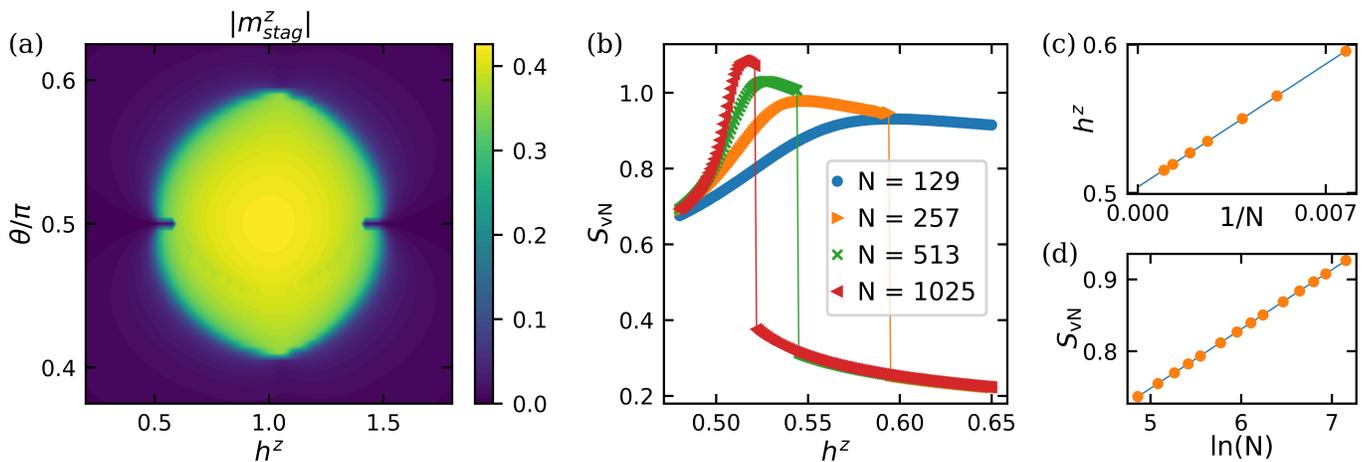}
    \caption{\label{fig:IsingCircle}Staggered magnetization per site $m^z_{stag}$ and entanglement entropy $S_{\rm{vN}}$ for the Ising ring. (a) The magnitude of the staggered magnetization per site $|m^z_{stag}|$ along the crystal $z$--axis as a function of $\theta$ and $h^z$ is plotted for $N=1025$. (b) The entanglement entropy $S_{\rm{vN}}$ as a function of $h^z$ at $\theta = \pi/2$ is plotted for different system sizes. (c) The peak positions of $S_{\rm{vN}}$ as a function of $1/N$ are fit to a polynomial to extrapolate the critical value of $h^z$ for the Ising transition at $\theta = \pi/2$ on the left side of the ring. (d) The entanglement entropy at the critical point found in (c) v.s. $\ln(N)$ is fit to the CFT form in Eq.~\eqref{eq:eescaling} and the extracted central charge $c = 0.498(3)$.}
  \end{center}
\end{figure*}

\subsection{Ising ring}
\label{subsec:isingcircle}

For $\theta=\pi/2$, the spin chain Hamiltonian Eq.~(\ref{eq:rotatedHam}) reads
\begin{eqnarray}
\label{eq:hamthetapi2}
\nonumber H_{\theta=\frac{\pi}{2}} &=& \sum_i \left( -\frac{1}{2}\sigma^x_i \sigma^x_{i+1} - \sigma^y_i \sigma^y_{i+1} + 2 \sigma^z_{i}\sigma^z_{i+1} \right) \\ &+& \left(h^z - 1\right)\sum_i \sigma^x_{i}.
\end{eqnarray}
Here, $J_z = 2|J_y| = 4|J_x|=2$, $\Gamma_x=1-h^z$, and $\Gamma_z = 0$. The dominant part of the Hamiltonian is a transverse-field Ising model $H_{Ising} = \sum_i \left[2 \sigma^z_{i}\sigma^z_{i+1} - \Gamma_x \sigma^x_i\right]$, which has a quantum phase transition belonging to the Ising universality class. A $\pi$-rotation about $z$-axis only reverses the direction of the $\Gamma_x$ field, so there should be two Ising critical points equidistant from $h^z=1$. The critical points of $H_{Ising}$ are given by $|h^z - 1| = 2$ \cite{R_B_Stinchcombe_1973}. However, the FM coupling $-\sigma^x\sigma^x/2$ will enhance the effects of transverse field, so the expected critical values follow $|h^z_c - 1| < 2$. At a deviation $\delta\theta$ from $\theta=\pi/2$ e.g., $\theta=\pi/2-\delta\theta$, one can show that in Eq.~\eqref{eq:rotatedHam}, $J_x$, $J_z$, and $\Gamma_x$ all deviate from the values at $\theta=\pi/2$ by order of $(\delta\theta)^2$, so the transverse-field Ising model persists to be the dominant part of the Hamiltonian as long as $\delta\theta$ is small. As $\theta$ is decreased from $\pi/2$, the positive $J_z$ decreases, the negative $J_x$ decreases, and $|\Gamma_z|$ increases, so the strength of the $\Gamma_x$ field at the critical points should decrease until a tiny transverse field can induce an Ising transition. Therefore, it is expected to have a ring of Ising critical points centered at $\theta=\pi/2$ and $h^z=1$, up to a correction of order of $(\delta\theta)^2$.

We do see a ring of critical points in the phase diagram shown in Fig.~\ref{fig:NVphasediagramspinprofile}(b). To demonstrate that the phase transitions belong to the Ising universality class, we show the staggered magnetization per site $|m^z_{stag}|$ in Fig.~\ref{fig:IsingCircle}(a), where there is a circular area in which $|m^z_{stag}| > 0$ and out of which $|m^z_{stag}| = 0$. This observation indicates an AFM phase inside the ring and a quantum phase transition associated with $\mathbb{Z}_2$ symmetry breaking. Moreover, this region can be verified as belonging to the Ising universality class by showing that the central charge $c=1/2$. To find $c$ for the transition, it is necessary to find the critical field $h^z_c$ where the Ising transition occurs. We first plot $S_{\rm{vN}}$ as a function of $h^z$ for different system sizes, $N$, shown in Fig.~\ref{fig:IsingCircle}(b). Interpolations are performed to accurately locate the peak positions of $S_{\rm{vN}}$. Subsequently, in Fig.~\ref{fig:IsingCircle}(c) we plot the corresponding values of $h^z$ for $S_{\rm{vN}}$ peaks as a function of $1/N$. We then fit the peak positions with a high degree polynomial of $1/N$ to obtain the value of the critical field $h^z_c$ in the thermodynamic limit $N\rightarrow \infty$. The extrapolated value of $h^z_c$ is $0.5036(2)$ and the corresponding $|\Gamma_x| = 0.4964(2) < 2$ as expected.  We finally calculate the entanglement entropy for different $N$ at $h^z_c$, and fit the data with the expected scaling predicted by CFT and given by Eq.~(\ref{eq:eescaling}). The results are presented in Fig.~\ref{fig:IsingCircle}(d), and yield the fitted value of $c = 0.498(3)$, which is consistent with the value $c=0.5$ for Ising CFT. In Appendix~\ref{appdx:IsingRing} we show results for $\theta = 1.4 \approx 0.45\pi$, where we found $h^z_c = 0.6298(12)$ and $c = 0.50(3)$. All of these observations verify that the critical ring is of Ising universality class.

\subsection{$\Gamma_x=0$: Incommensurate line}
\label{subsec:gammax0}
\begin{figure*}[t!]
  \begin{center}
    \includegraphics[width=\textwidth]{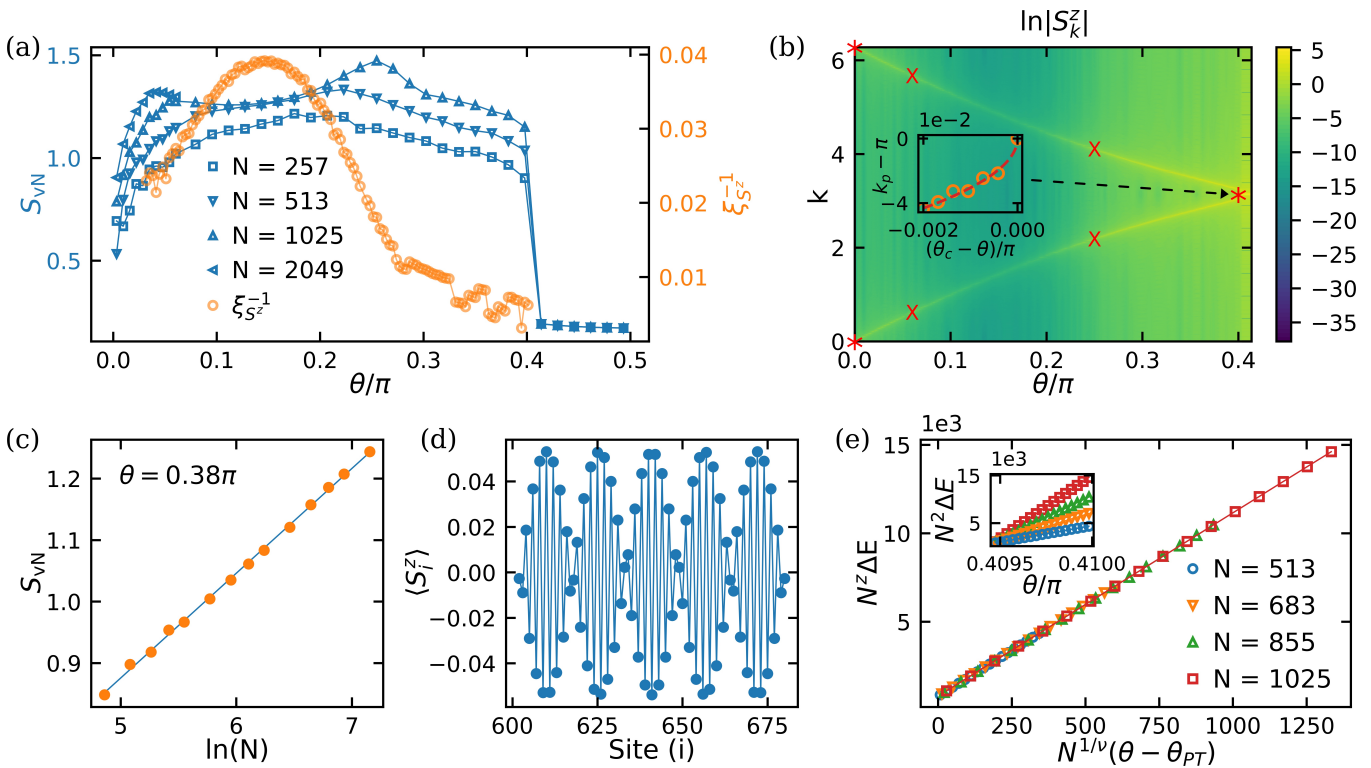}
    \caption{\label{fig:Gx=0}Critical properties along the $\Gamma_x = 0$ line. (a) The entanglement entropy $S_{\rm{vN}}$ as a function of $\theta$ is plotted for different system sizes on the left $y$ axis. The results for $N=2049$ sites only contain data with $\theta \leq 0.064\pi$. The inverse correlation length $\xi^{-1}_{S^z}$ extracted from connected correlators $C(r) = \langle S^z_{i_0} S^z_{i_0+r} \rangle - \langle S^z_{i_0}\rangle \langle S^z_{i_0+r}\rangle$ with $i_0 = (N-1)/2$ is plotted on the right $y$ axis. (b) The filled contour plot shows the absolute value of the discrete Fourier Transform of the spin density profile $|\tilde{S}^z_k|$ as a function of $\theta$ and $k$ on a natural logarithmic scale for $N=1025$. The peak position $k_p$ of $|\tilde{S}^z_k|$ changes continuously from $k_p=0$ ($k_p=2\pi$) to $k_p=\pi$ with $\theta$. The three red asterisks show the locations of PT points, while the four red X's show the locations of BKT points. The inset shows $k_p$ v.s. $\theta$ fit to their power-law near the PT point with $k_p=\pi$. (c) The entanglement entropy at $\theta = 0.38\pi$ as a function of $\ln{N}$ is fit to the CFT form in Eq.~\eqref{eq:eescaling} to find the central charge $c = 1.02(3)$. (d) The spin density profile at $\theta = 0.38\pi$ is plotted for $N = 1281$. (e) The best data collapse of the rescaled energy gap $N^z\Delta E$ v.s. $N^{1/\nu}(\theta-\theta_{PT})$ for various system sizes is presented at the optimal values of the PT point $\theta_{PT}$ and critical exponents $\nu$ and $z$. The inset shows the rescaled energy gap $N^2\Delta E$ as a function of $\theta/\pi$.}
  \end{center}
\end{figure*}
We have argued in Sec.~\ref{subsec:spinhalfH} that the $\Gamma_x = 0$ line is special, while the $\Gamma_z=0$ line is trivial without any criticality, so the $\Gamma_x=0$ line will be analyzed in this section. Notice that any field perturbation along the $x$ direction, i.e., $\Gamma_x=0\rightarrow\delta\Gamma_x$, will reduce the symmetry of the XYZ model in the presence of an external field along the $z$-direction [See Eq.~(\ref{eq:rotatedHam})]. Both spontaneous and explicit global symmetry breakings result in energetic domain wall excitations and create a gapped phase with low entanglement between subsystems. Thus, the entanglement entropy peaks exactly at $\Gamma_x = 0$, as shown in Fig.~\ref{fig:NVphasediagramspinprofile}(b), and any criticality must be strictly on $\Gamma_x=0$ line without finite width. 

In the phase diagram Fig.~\ref{fig:NVphasediagramspinprofile}(b), it is observed that the $\Gamma_x=0$ line connects the circular phase boundary to the PT point at $\theta = 0$. As discussed in Sec.~\ref{subsec:spinhalfH}, the competition between an AFM coupling and the tendency to align along the external field in the presence of transverse FM couplings may cause proliferation of domain wall excitations and create a floating phase. The floating phase is an incommensurate density wave order with gapless excitations and emergent U(1) symmetry \cite{VillainBak1981,PhysRevLett.49.793,GiamarchiQP1D2003}. On the $\Gamma_x=0$ line below the ring, we indeed observe an incommensurate density wave at $\theta=0.4\pi$ [See the inset of Fig.~\ref{fig:NVphasediagramspinprofile}(b)], where the wavelength is fractional in units of lattice spacing but close to $2$, due to proximity to the AFM phase which has a periodicity of $2$. The big envelope is due to the superposition of two waves with close wave vectors: one is the floating phase and the other is the intrinsic wave defined by the lattice. Another example of an incommensurate density wave is shown in Fig.~\ref{fig:Gx=0}(d) for $\theta=0.38\pi$, where the wave length deviates more from $2$ and the envelope is smaller. These incommensurate density waves with varying wavelength show that we have a floating phase on the $\Gamma_x=0$ line.

Based on the theory of TLL \cite{F.D.M.Haldane_1981}, a large class of 1d critical phases is described by a free boson field with a renormalized stiffness $K$ (Luttinger liquid parameter), where the bosons act as density fluctuations and propagate with an effective velocity $v$. The PT transition between 1d critical phases and crystalline orders happens when this velocity becomes zero, and the physical system has a spectrum where the leading dispersion relation is quadratic with the momentum of low-energy excitations \cite{MilaLifshitz2021}. Additionally, the BKT transition happens when the Luttinger-liquid parameter reaches the value at which TLL is unstable and becomes a disordered phase with exponentially decaying correlations \cite{GiamarchiQP1D2003,F.D.M.Haldane_1981}. The values of $K$ and $v$ in the effective TLL theory change continuously with coupling constants in the physical system. This suggests there is a critical floating phase near the PT point at $\theta=0$, and the $\Gamma_x=0$ line must have another PT point between the floating phase and AFM phase. Our model is similar to the XY model in an external field along the $x$-direction, which is dual to the quantum ANNNI model and has a critical floating phase bounded by a BKT line and a PT line \cite{PhysRevB.67.094435}. As a result there are likely BKT transitions between the floating phase and the disordered phase on the $\Gamma_x=0$ line, and this is investigated further in this section.

To numerically investigate our predictions, we plot the entanglement entropy $S_{\rm{vN}}$ as a function of $\theta$ for $N=257$, $513$, $1025$, and $2049$ in Fig.~\ref{fig:Gx=0}(a). We only provide data with $\theta \le 0.064\pi$ for $N=2049$ to confirm criticality for small $\theta$. It is seen that the entanglement entropy around $0.15\pi$ saturates at $N \gtrsim 512$, indicating that there exists a noncritical gapped phase on the $\Gamma_x=0$ line around this point. Near $\theta=0$ and $\theta=0.41\pi$, $S_{\rm{vN}}$ continues to increase with the system size for large $N$, confirming that there exist two separate critical floating phases. There are two BKT transition points on the $\Gamma_x=0$ line between the two floating phases and the gapped phase, which are signaled by the peaks of $S_{\rm{vN}}$ at $\theta \approx 0.04\pi$ and $\theta \approx 0.26\pi$. We also extract the correlation length $\xi_{S^z}$ from the connected correlators $C(r) = \langle S^z_{i_0} S^z_{i_0+r}\rangle - \langle S^z_{i_0}\rangle \langle S^z_{i_0+r}\rangle$ with $i_0=(N-1)/2$ for $N=1025$ \cite{PhysRevLett.122.017205} and plot $\xi^{-1}_{S^z}$ as a function of $\theta$ in Fig.~\ref{fig:Gx=0}(a). One can see that $\xi^{-1}_{S^z}$ is maximized around $0.14\pi$ and decreases towards $\theta=0$ and $\theta=0.41\pi$, which is consistent with the two floating phases having divergent correlation lengths, while the correlation length of the gapped phase between them is finite~\cite{PhysRevB.90.214426}.

The PT point at $\theta=0$, has a spin density profile $\langle S^z_i \rangle$ that is flat, which corresponds to a density wave vector $k_p = 0$, while the PT point at $\theta \approx 0.41\pi$ on the AFM phase boundary has staggered magnetization ordering, which corresponds to $k_p = \pi$. We use a natural logarithmic scale to show the absolute value of the discrete Fourier transform of the spin density profile with the magnetization subtracted $|\tilde{S}^z_k| = |\sum_n{(S^z_n-m^z)\exp{(-ikn)}}|$ on the $\Gamma_x=0$ line, and as a function of $\theta$ and the wave vector $k$ for $N=1025$ in Fig.~\ref{fig:Gx=0}(b). When calculating $|\tilde{S}^z_k|$ we used sites indexed between $N/3$ and $2N/3$ in $m^z$ and the sum over $n$ in order to remove some edge effects. The peak position of $k$ gives the density wave vector $k_p$, which characterizes the main oscillation pattern in the spin density waves. It is clear that there exist two smooth lines of peaks of $|\tilde{S}^z_k|$, symmetric about $k=\pi$, connecting $k_p = 0$ ($2\pi$) at $\theta=0$ and $k_p=\pi$ at $\theta\approx 0.41\pi$. Thus the density wave vector $k_p$ changes continuously from $0$ ($2\pi$) to $\pi$ as $\theta$ is tuned from $0$ to about $0.41\pi$. It is also seen that the peak height of $|\tilde{S}^z_k|$ decreases with $\theta$, until it reaches a minimum near $\theta=0.15\pi$ where it begins to increase. The floating phase has an incommensurate density wave order and the peak height of $|\tilde{S}^z_k|$ should be finite, while the disordered phase has no density wave order in the thermodynamic limit and should have weaker oscillations in spin density profile than the floating phase in finite-size systems. The results in Fig.~\ref{fig:Gx=0}(b) are consistent with there being a gapped disordered phase between two floating phases on the $\Gamma_x=0$ line.

To further quantify the criticality of the floating phase, we study the scaling of the entanglement entropy at $\theta = 0.38\pi$ on $\Gamma_x=0$ line. The spin density profile is plotted in Fig.~\ref{fig:Gx=0}(d) and one can see it is indeed an incommensurate density wave. The entanglement entropy as a function of $\ln{N}$ is plotted in Fig.~\ref{fig:Gx=0}(c), where we find $S_{\rm{vN}}$ is grows linearly with $\ln{N}$. We fit the data for $S_{\rm{vN}}$ to the CFT form in Eq.~\eqref{eq:eescaling} and obtain the central charge $c = 1.02(3)$, which is consistent with the theory of TLL or Gaussian CFT with $c=1$ \cite{Nomura_1998}.

We finally provide numerical evidence for the PT transition between the floating phase and AFM phase on the $\Gamma_x=0$ line. Notice that the entanglement entropy suddenly drops as $\theta$ increases past $\theta \approx 0.41\pi$ where the $\Gamma_x = 0$ line crosses the AFM ring, which clearly distinguishes the PT point from a BKT point where $S_{\rm{vN}}$ changes smoothly across the transition. The critical exponents for the PT transitions are $\bar{\beta} = \nu = 1/2$ and $z=2$. The $\bar{\beta}$ exponent describes the power-law behavior of the density wave vector $k_p-\pi \sim (\theta_c - \theta)^{\bar{\beta}}$ for the floating phase near the PT point. We fit the peaks of $|\tilde{S}^z_k|$ in Fig.~\ref{fig:Gx=0}(b) to a power-law form and obtain $\bar{\beta} = 0.50(2)$, consistent with the expected value $1/2$. On the other hand, when approaching the PT point from the gapped side, the correlation length diverges as $\xi \sim (\theta-\theta_c)^{-\nu}$ and the energy gap closes as $\Delta E \sim 1/N^z$. A scaling hypothesis can be postulated for the energy gap around the critical point $N^z\Delta E = f(N^{1/\nu}[\theta-\theta_c])$, where $f(x)$ is a universal function of $x$. We calculate the energy gap for $N=513$, $683$, $855$, and $1025$ at values of $\theta$ close to the phase transition point but inside the AFM phase and plot $N^2\Delta E$ v.s. $\theta/\pi$ in the inset of Fig.~\ref{fig:Gx=0}(e). There exists a fixed crossing point for $N^2\Delta E$ near $\theta = 0.4095\pi$, indicating a phase transition there. We then fit $N^z\Delta E$ as a function of $N^{1/\nu}(\theta-\theta_{\rm{PT}})$ to a high-degree polynomial, where $\theta_{\rm{PT}}$, $\nu$, and $z$ are tuning parameters. By minimizing the mean squared residuals for the curve fit, the location of PT point $\theta_{PT}$ and the critical exponents $z$ and $\nu$ can be determined. Figure.~\ref{fig:Gx=0}(e) shows the optimal results for the curve fit where $\theta_{PT}=0.40946260(17)\pi$, $z=1.963(33)$, and $\nu=.5093(77)$. These results are consistent with the expected values $\nu=1/2$ and $z=2$ for PT transitions.

In summary, we have provided strong evidence in this section that on the $\Gamma_x=0$ line, there exist two separate critical floating phases each bounded by a PT point and BKT point, between which is a gapped disordered phase bounded by two BKT points.

\section{Conclusions} \label{sec:conclusion}
We propose a novel solid-state quantum simulator based on a 1D chain of spin centers implanted in SiC or diamond. We show that by considering the magnetic dipole-dipole interaction between $S=1$ spin centers, and an applied magnetic field, we are able to obtain an effective $S=1/2$ interacting spin chain defining our quantum simulator. Most importantly, we show that the corresponding effective Hamiltonian can be tuned with different values of magnetic field, and with the angle $\theta$ between the direction of the spin center array and the spin center main symmetry axis.
These enable our quantum simulator to be mapped to both isotropic Heisenberg model in the presence of a longitudinal field and spin chains in the universality class of the transverse-field Ising model. Furthermore, between these regimes, we find a line of enhanced entanglement entropy that presents a number of interesting behaviors, namely critical floating phases characterized by incommensurate spin density waves, and associated Pokrovsky-Talapov and Berezinskii-Kosterlitz-Thouless transitions. There has been much interest in realizing floating phases and studying commensurate-to-incommensurate phase transitions \cite{PhysRevLett.122.017205,PhysRevB.80.245418,PhysRevResearch.4.013093}, both with corresponding experiments based on cold atoms \cite{PhysRevLett.90.130401} and experimental proposals using Rydberg-atom arrays~\cite{chepiga_natcomm}. This is the first proposal of a quantum simulator based on spin centers in solid-state materials for realizing floating phases, where PT transitions between the commensurate AFM phase and incommensurate floating phases can be probed experimentally.

\vskip20pt
\begin{acknowledgments}
This work was supported in part by the National Science Foundation (NSF) RAISE-TAQS under Award Number 1839153 (S.W.T.), by the U.S. Department of Energy (DOE), Office of Basic Energy Sciences under Award Number DE-SC0019250 (M. E. F.) for the NV Hamiltonian derivation and DE-SC0019139 (Y.M.) for using quantum spin chains as quantum simulators. J.Z. is supported by NSFC under Grants No. 12304172 and No. 12347101, Chongqing Natural Science Foundation under Grant No. CSTB2023NSCQ-MSX0048, and Fundamental Research Funds for the Central Universities under Projects No. 2023CDJXY-048 and No. 2020CDJQY-Z003. Computations were performed using the computer clusters and data storage resources of the UCR High Performance Computing Center (HPCC), which were funded by grants from NSF (MRI-1429826) and NIH (1S10OD016290-01A1).
\end{acknowledgments}

\appendix

\section{The effective spin--1/2 Hamiltonian from spin centers} \label{appdx:effspinhalf}

In the NV$^-$ centers, if the levels $\ket{-1}$ and $\ket{0}$ are nearly degenerate, to the first-order approximation, we just need to keep the submatrix elements of spin--1 operators associated with the two states, then $S^x_{S=1} \rightarrow \sqrt{2}S^x_{S=1/2}$, $S^y_{S=1} \rightarrow \sqrt{2}S^y_{S=1/2}$, and $S^z_{S=1} \rightarrow S^z_{S=1/2}-1/2$. The effective spin-$1/2$ Hamiltonian is
\begin{eqnarray}
\label{eq:effhamoriginal}
\nonumber \frac{H}{J} &=& \sum_{i=1}^{N-1} \bigg\{ \left[3\sin^2(\theta) - 1\right]S_i^x S_{i+1}^x - S^y_i S^y_{i+1} \\ \nonumber &+& \frac{3\cos^2(\theta) - 1}{2}\left(S^z_i S^z_{i+1} + \frac{1}{4}\right) \\ \nonumber &+& \frac{3\sin(2\theta)}{2\sqrt{2}}\left(S^x_i S^z_{i+1} + S^z_i S^x_{i+1}\right)\bigg\} \\ \nonumber &-& \sum_{i=1}^{N}\left[\frac{h^z_{ex} +3\cos^2(\theta) - 1}{2}S^z_i + \frac{3\sin(2\theta)}{2\sqrt{2}}S^x_i + \frac{h^z_{ex}}{4}\right] \\ \nonumber &+& \left[\frac{3\cos^2(\theta) - 1}{4}(S^z_1 + S^z_N) + \frac{3\sin(2\theta)}{4\sqrt{2}}(S^x_1 + S^x_N)\right] \\
\end{eqnarray}
After leaving out the boundary terms and the constants, which do not change the criticality, and replacing the spin-$1/2$ operators by Pauli matrices, we obtain Eq.~(\ref{eq:effectiveham}).

\begin{figure*}[t!]
  \begin{center}
    \includegraphics[width=\textwidth]{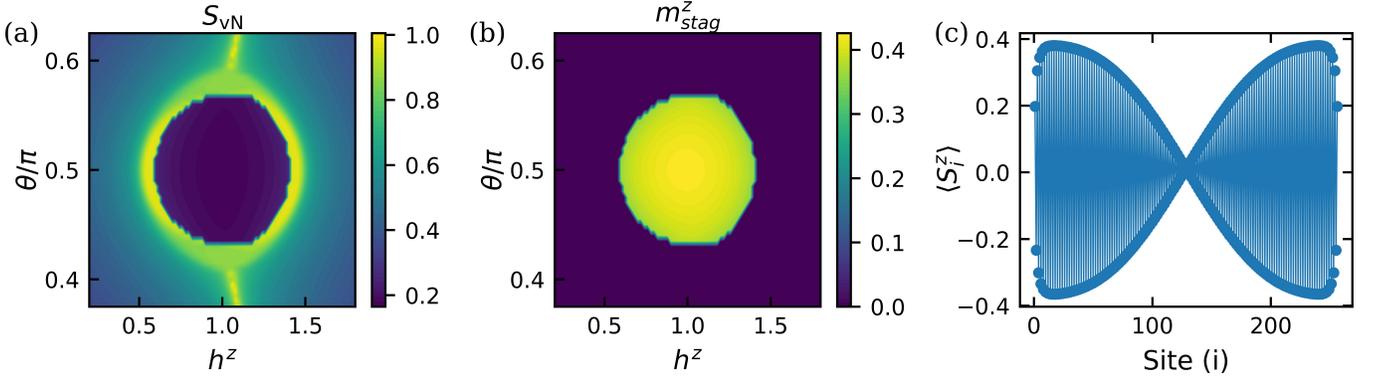}
    \caption{\label{fig:evensiteseffects}Results for even number of sites $N=256$. (a) The filled contour plot shows the entanglement entropy $S_{\rm{vN}}$ in the $h^z$--$\theta$ plane. (b) The filled contour plot shows the staggered magnetization per site along the crystal z-axis $m^z_{stag}$. (c) The spin density profile is plotted at $\theta=0.42\pi$ and $h^z = 1.0$ on the yellow entanglement plateau in (a). The domain wall forms due to boundary effects with even number of sites.}
  \end{center}
\end{figure*}

\section{Rotation of the spin-$1/2$ Hamiltonian}
\label{appd2}

Rotating the Hamiltonian Eq.~(\ref{eq:rotatedHam}) around the $y$-axis by an angle $\alpha$, the new spin operators change as follows
\begin{eqnarray}
    \sigma^x \rightarrow e^{is^y\alpha}\sigma^xe^{-is^y\alpha} = \cos(\alpha)\sigma^x + \sin(\alpha)\sigma^z, \\
    \sigma^z \rightarrow e^{is^y\alpha}\sigma^ze^{-is^y\alpha} = \cos(\alpha)\sigma^z - \sin(\alpha)\sigma^x.
\end{eqnarray}
Then the new Hamiltonian can be written as
\begin{eqnarray}
\nonumber \tilde{H} &=& \sum_i \left( J_{x} \sigma^x_i \sigma^x_{i+1} + J_{z} \sigma^z_{i} \sigma^z_{i+1} -\sigma^y_i \sigma^y_{i+1}\right) \\ \nonumber &+& \sum_i J_{xz} \left(\sigma^x_i \sigma^z_{i+1} + \sigma^z_i \sigma^x_{i+1}\right) \\ &-& \sum_i \left( \Gamma_z\sigma^z_i + \Gamma_x\sigma^x_i \right),
\end{eqnarray}
where
\begin{eqnarray}
J_{x} &=& \frac{A-B}{2}\cos\left(2\alpha\right) - C\sin\left(2\alpha\right) + \frac{A+B}{2}, \\ 
J_{z} &=& -\frac{A-B}{2}\cos\left(2\alpha\right) + C\sin\left(2\alpha\right) + \frac{A+B}{2}, \\
J_{xz} &=& \frac{A-B}{2}\sin(2\alpha) + C\cos(2\alpha), \\
\Gamma_z &=& \left(h^z_{ex}+2B\right)\cos(\alpha) + 2C\sin(\alpha), \\
\Gamma_x &=& -\left(h^z_{ex}+2B\right)\sin(\alpha) + 2C\cos(\alpha),
\end{eqnarray}
for $A$, $B$, $C$, and $\alpha$ defined in Sec.~\ref{subsec:spinhalfH}.

Let $J_{xz} = 0$, we can solve for $\alpha$ such that the $\sigma^x_i\sigma^z_{i+1}+\sigma^z_i\sigma^x_{i+1}$ terms are eliminated. The solution is given by $\tan(2\alpha) = 2C/(B-A)$ if $\theta \neq \arcsin(2/3)$ or $\alpha=\pi/4$ if $\theta = \arcsin(2/3)$, where $\theta = \arcsin(2/3)$ is the angle for $A = B$. Notice that the $\arctan$ function returns values between $-\pi/2$ and $\pi/2$, so there is a discontinuity in $\alpha$ when $B-A$ becomes negative from positive as we increase $\theta$. We can let $\alpha \in [0, \pi/2]$, then we obtain an expression of $\alpha$ as a continuous function of $\theta$ in Eq.~(\ref{eq:alphavstheta}).

\section{DMRG Specifications}
\label{appdx:DMRG}

 Our DMRG calculations are performed with \textsc{ITensor Julia Library} \cite{10.21468/SciPostPhysCodeb.4}.  When searching for the ground state, we gradually increase the maximum bond dimension $D$ during the variational sweeps until the truncation error $\epsilon$ is below $10^{-10}$. Some high-precision calculations have truncation errors between $10^{-11}$ and $10^{-12}$ to ensure that the largest bond dimension of the sweeps reaches $\gtrsim 100$. DMRG sweeps are terminated once the ground-state energy changes less than $10^{-11}$ and the von Neumann entanglement entropy changes less than $10^{-8}$ between the last two sweeps. In this work, the largest bond dimension in the final sweeps for $\epsilon=10^{-10}$ is about $D = 350$ for $N = 1025$, $\theta = 0.27\pi$, and $h^z = h_c$, where $h^z_c$ is a solution for the $\Gamma_x = 0$ line. All results shown use open boundary conditions (OBCs).

\section{Effects of odd and even number of sites}
\label{appdx:EvenVsOdd}

On the boundary sites, because the interaction only comes from one side in the bulk, they often prefer some particular states favored by the external field. Near the AFM transition at the ring in our phase diagram, if we have even number of sites, the AFM state will have opposite spin states on the boundary sites. This is not favored by a uniform external field, so a state with a domain wall in the center may exist near the phase transition. Fig.~\ref{fig:evensiteseffects}(a) shows that there exists a high entanglement plateau before the transition into the AFM phase, which is due to the domain-wall state that introduces a large entanglement constant. The constant will not scale with the system size, so it will not change the critical point. This high plateau may overwhelm the entanglement scaling at small system sizes, so we use odd number of sites to remove the domain-wall state. In the domain-wall state, there is a reflection symmetry, each half of the chain has a non-zero staggered order, but the staggered order is zero for the whole system. One can see from Fig.~\ref{fig:evensiteseffects}(b) that the circle of non-zero staggered order is smaller than the one for odd number of sites $N=257$ shown in Fig.~\ref{fig:IsingCircle}(a). Fig.~\ref{fig:evensiteseffects}(c) shows the AFM state with a domain wall at $\theta=0.42\pi$ and $h^z = 1.0$ on the yellow entanglement plateau.

\section{Additional data for Ising ring}
\label{appdx:IsingRing}

\begin{figure*}[t!]
  \begin{center}
    \includegraphics[width=\textwidth]{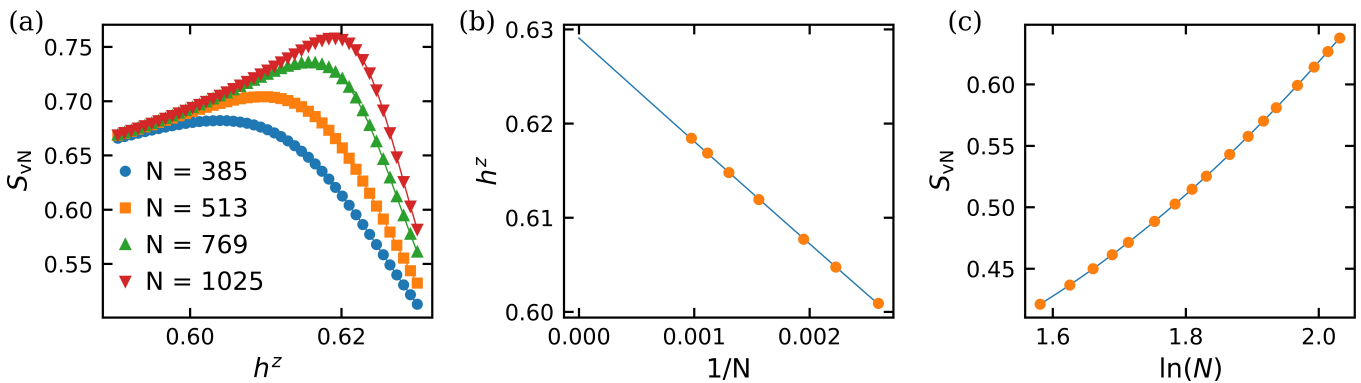}
    \caption{\label{fig:theta=1.4}Entanglement entropy scaling for the Ising ring when $\theta = 1.4$. (a) The entanglement entropy $S_{\rm{vN}}$ plotted as a function of $h^z$ for different system sizes. (b) The peak positions of $S_{\rm{vN}}$ v.s. $1/N$ are extrapolated to the thermodynamic limit where $1/N = 0$. (c) $S_{\rm{vN}}$ as a function of $\ln(N)$ is fit to Eq.~\eqref{eq:S1ScalingCorrections} to extract the central charge $c = 0.50(3)$.}
  \end{center}
\end{figure*}

This section shows data and the process used to confirm that the phase transition at $\theta = 1.4 \approx 0.45\pi$ belongs to the Ising universality class. Similarly to Section~\ref{subsec:isingcircle}, in Fig.~\ref{fig:theta=1.4}(a) we show the entanglement entropy $S_{\rm{vN}}$ for several system sizes $N$, as $h^z$ is tuned across the phase transition. This clearly shows that for different system sizes, the entanglement entropy peaks for different values of $h^z$, which indicates that the phase transition occurs at different $h^z$. For each system size we use interpolation to precisely find $h^z$ at the peak and plot these values of $h^z$ v.s. $1/N$ in Fig.~\ref{fig:theta=1.4}(b), where the data is fit to a high degree polynomial and extrapolated to $1/N = 0$ to find that the phase transition occurs at $h^z_c = 0.6298(12)$ in the thermodynamic limit. Eq.~\eqref{eq:eescaling} from CFT predicts that at a critical point such as this, the entanglement entropy should scale linearly with $\ln(N)$ in the thermodynamic limit. However, finite-size effects cause this to be a poor fit for our data in Fig.~\ref{fig:theta=1.4}(c), which shows the scaling of entanglement entropy $S_{\rm{vN}}$ as a function of $\ln{(N)}$. This is corrected by adding a term to the fit function that goes to $0$ as $N$ goes to infinity, but greatly improves the quality of the fit. The data in Fig.~\ref{fig:theta=1.4}(c) is fit to
\begin{eqnarray}
\label{eq:S1ScalingCorrections}
    S_{\rm{vN}} = s_0 + \frac{c}{6}\ln (N) + Ke^{-N},
\end{eqnarray}
for the non-universal constants $s_0$ and $K$, to find the central charge $c = 0.50(3)$, which confirms that this transition and all transition on the Ising ring are of the Ising universality class where $c = 1/2$.

\providecommand{\noopsort}[1]{}\providecommand{\singleletter}[1]{#1}%

\end{document}